\documentclass{ws-procs10x7}
\usepackage{balance}

\newcolumntype{d}[1]{D{.}{.}{#1}}

\def\Journal#1#2#3#4{{\it #1} {\bf #2}, #3 (#4)}

\makeindex
\begin{document}

\title{New Results on $X(3872)$ from CDF}

\author{M.~Kreps$^*$ on behalf of the CDF Collaboration}

\address{Institut f\"ur Experimentelle Kernphysik, University
of Karlsruhe, Germany \\$^*$E-mail: kreps@ekp.uni-karlsruhe.de}


\twocolumn[\maketitle\abstract{
In 2003 the X(3872) particle was discovered by the Belle collaboration.
Despite results collected since then, the nature of the state still
remains unclear. In this contribution we report on new results on
properties of the X(3872) state using data collected with
CDF II 
detector at the Fermilab Tevatron. The dipion mass
spectrum and angular distributions are used to determine the 
$J^{PC}$ quantum numbers of the state.}
\keywords{$X(3872)$; charmonium; exotic mesons.}
]

\section{Introduction}

The recent discovery of the $X(3872)$ state\cite{belleDiscovery,cdfDiscovery} 
led to new interest in
charmonium spectroscopy. It was found in a search for
missing charmonium resonances. Despite enormous experimental
and theoretical effort its exact nature is still
unknown. Shortcomings of conventional explanations and
the proximity of the $D^0D^{*0}$ threshold have raised questions
whether the $X(3872)$ could be an exotic form of matter, e.g. a
mesonic molecule, $c\overline{c}g$ hybrid, etc.

Important for the understanding of the 
$X(3872)$ state are the quantum numbers spin $J$, parity $P$
and charge conjugation parity $C$. Here we present a determination of
these quantum numbers using $X(3872)\rightarrow J/\psi
\pi\pi$ decays collected by the CDF II detector\cite{cdfDetector}.
As quantities sensitive to the $J^{PC}$ quantum numbers, the dipion invariant mass and
the angular distributions are used. In both cases, the
measured distributions 
are compared to the predictions for different
$J^{PC}$ hypothesis to infer the $J^{PC}$ quantum numbers.

Details of the presented analysis can be 
found in Ref.\cite{cdfDipion,CDFJPC}.

\section{Theoretical predictions}

To obtain predictions for different $J^{PC}$ combinations, we
consider the $X(3872)$ decay as a sequence of  two-body
decays. First, the $X(3872)$ decays to the $J/\psi$ and
$(\pi\pi)_{s,p}$ system, which is followed by decays
$J/\psi\rightarrow \mu^+\mu^-$ and $(\pi\pi)_{s,p}\rightarrow
\pi^+\pi^-$. In this view, the full decay amplitude is
given by three amplitudes, one for each of the two-body
decays and two "propagators", which connect vertices of the
two-body decays. The amplitudes for the decay vertices are
obtained using the helicity formalism. 

The final amplitude for a given $J^{PC}$ hypothesis is
obtained from the squared total matrix element by averaging over
all initial state helicities, incoherently summing over all
final state helicities and coherently summing over
intermediate state helicities.

The angular distributions
are determined fully by the matrix elements for the vertices, which
for the fixed helicities are given by the Wigner functions.
In general more than one possibility to form spin $J$ from
the relative angular momentum $L$ and spin $S$ of the daughter
particles exist. Of these independent amplitudes, only
the ones with lowest $L$ are used, as the higher $L$ amplitudes are
usually suppressed. 

The "propagator" for the $J/\psi$ is modelled by a $\delta$
function due to the very small width of the $J/\psi$.
For the $(\pi\pi)$ system "propagator", the situation is more complicated as
any description has some unknowns inside, which doesn't
allow one to make a precise prediction. For the $(\pi\pi)$ system
in the S-wave we use multipole expansion\cite{multipoleExp}. The $(\pi\pi)$
system in P-wave is described by the intermediate $\rho$
resonance for which we use a relativistic Breit-Wigner formula
\begin{equation}
\frac{d\Gamma_X}{dm_{\pi\pi}}\,=\,2m_{\pi\pi}
\frac{\Gamma_{X}(m_{\pi\pi})\cdot
2m_\rho\Gamma_{\rho}(m_{\pi\pi})}
       {(m_{\pi\pi}^2-m_\rho^2)^2+m_\rho^2\Gamma_\rho^2(m_{\pi\pi})}
 \nonumber
\end{equation} 
In a case of broad resonances such as $\rho$ the width in
the Breit-Wigner formula has to by modified to 
\begin{equation}
\Gamma_{\rho}\,=\,\Gamma_{0,\rho}\bigg(\frac{k^*}{k^*_0}\bigg)^{2L+1}
\bigg(\frac{f(k^*)}{f(k^*_0)}\bigg)^2
     \bigg(\frac{m}{m_0}\bigg)  \nonumber
\end{equation}
due to the variation of the kinematic
factor across the width. Here, $k^*$ is the momentum of the
decay product in the centre-of-mass system of the decaying particle
and $f(k^*)$ is a form-factor. The model suggested by
Blatt and Weisskopf is used in our analysis\cite{Blatt-Weisskopf}. 
This model has one free parameter,
which is the effective radius of the resonance for which typical
values are in the range from $0.3$ fm to $1$ fm\cite{Rvalues}. Another
complication arises from possible $\rho$-$\omega$
interference, which is also included in our description.

\section{Results}

\begin{figure}[b]
\centerline{\psfig{file=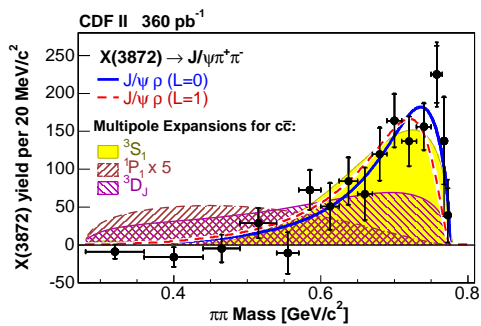,width=2.7in}}
\caption{Dipion invariant mass distribution with the predictions
from multipole expansions and $L=0$ and $L=1$ decay
to $J/\psi\,\rho$.}
\label{fig1}
\end{figure}

The first distribution, to which we look is dipion invariant
mass distribution\cite{cdfDipion}. It is shown in Figure
\ref{fig1}. The $^3S_1$, $^1P_1$ and $^3D_J$ multipole
expansion for charmonia and $L=0$ and $L=1$ decay to the
$J/\psi\,\rho$ were tested. Out of the tested models the $^3S_1$ multipole expansion
and both $L=0$ and $L=1$ decay to the $J/\psi\,\rho$ are
able to fit the data.
While $^3S_1$ multipole expansion is able to describe data,
it is disfavoured as this would be in tension with
non-observation of the $X(3872)$ by the BES
experiment\cite{BES}. Therefore only decays to
$J/\psi\,\rho$ remain as a viable options for the $X(3872)$
decay. Unfortunately, at the current level of
understanding, it is not possible 
 to distinguish between $L=0$ or
$L=1$ decays. The main reason is the 
uncertainty in the modelling of the dipion mass shape. With
reasonable parameters for the form-factors both can describe
data. If we allow in addition $\rho$-$\omega$ mixing,
for certain mixing phases, the $L=1$ describes data even better
than $L=0$. This conclusion is in contradiction to the
conclusion by Belle\cite{BelleJPC}. The origin
of this disagreement stems from different modelling of the
dipion mass distribution. The model used by the Belle
collaboration doesn't include the form-factor in the Breit-Wigner 
formula. If we drop the form-factor from the description,
the CDF dipion mass distribution is also inconsistent with the
$L=1$ decay to the $J/\psi\,\rho$.
As it is not clear which model is the correct
one and therefore, one should remain cautious at this stage.

\begin{figure}[b]
\centerline{\psfig{file=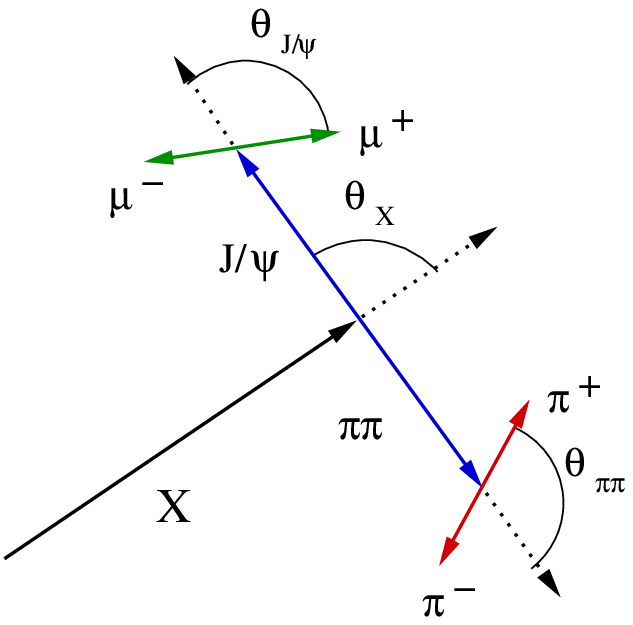,width=1.4in}}
\centerline{\psfig{file=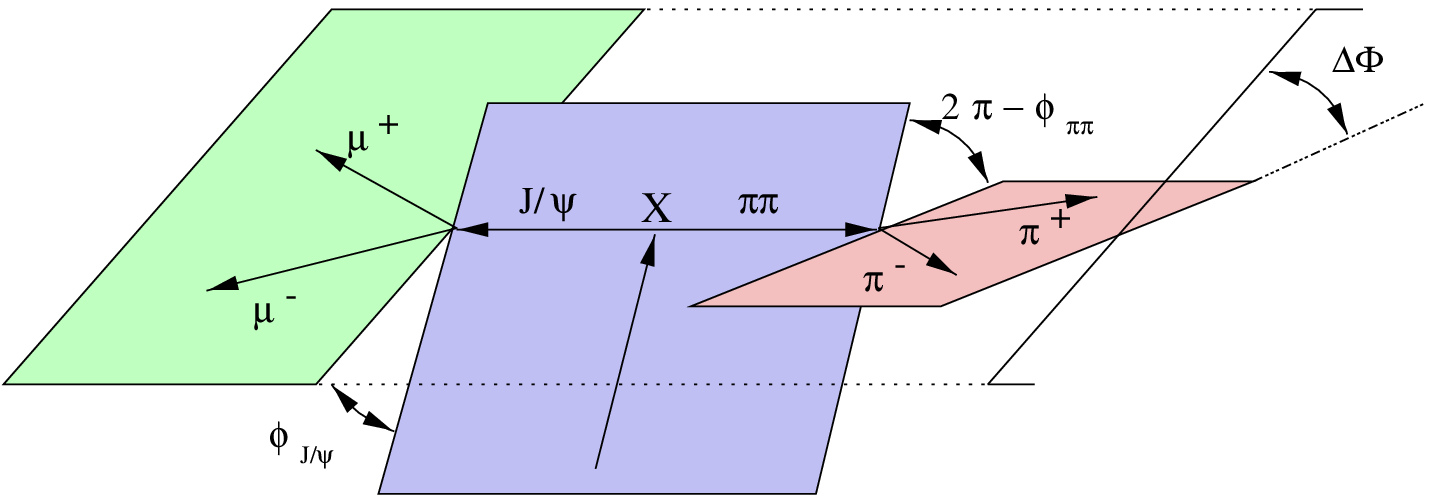,width=2.7in}}
\caption{Definition of the angles used in the angular
analysis.}
\label{fig2}
\end{figure}
In order to gain more information on the properties of the
$X(3872)$ we now consider angular
distributions\cite{CDFJPC}. The angles
describing the decay are defined in Figure \ref{fig2}. Out
of all angles, for unpolarised production, only three are
sensitive to the $J^{PC}$ quantum numbers. Those are
$\theta_{J/\psi}$, $\theta_{\pi\pi}$ and $\Delta\Phi$. The last
sensitive variable is the dipion invariant mass, but as we saw, there
is considerable ambiguity in modelling. 
Therefore to avoid wrong conclusions, we fix the dipion mass
distribution to an $L=0$ $\rho$ Breit-Wigner, which was found
to describe data.

\begin{figure}
\centerline{\psfig{file=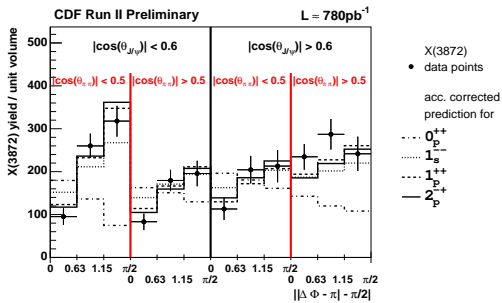,width=2.8in}}
\caption{Measured angular distributions (points) with
expectations for several $J^{PC}$ (lines).}
\label{fig3}
\end{figure}
To extract angular distributions from data, a slicing technique
with a binned maximum likelihood fit is used, where
the background is described by a second order polynomial and
the signal by a Gaussian. The position and width of the Gaussian
are fixed to the result of the fit to the full sample.
In order to increase the discriminating power of the analysis,
we exploit also correlations among the angles by usage of a
3-dimensional fit. We use $3\times 2\times 2$ binning, where
three bins are used for $\Delta\Phi$ angle. 
The measured
distributions are shown in Figure \ref{fig3} together with
the expectations for several  $J^{PC}$ hypotheses. To quantify
the agreement between data and expectations, a $\chi^2$
comparison is done. The resulting $\chi^2$ values for different
assignments are shown in Table \ref{tab1}. From the results we
conclude, that only the $1^{++}$ and $2^{-+}$ assignments are
able to describe data, while all the others are excluded by
more than $3\,\sigma$.
\begin{table}
\tbl{Result of the angular analysis for all tested
assignments.\label{tab1}}
{\begin{tabular}{@{}ccc@{}}
\toprule
 hypothesis &  3D $\chi^2$ / 11 d.o.f. & $\chi^2$ prob.\\
\colrule
 $1^{++}$  &   13.2         &   27.8\% \\
 $2^{-+}$  &   13.6         &   25.8\% \\
 $1^{--}$  &   35.1         &   0.02\% \\
 $2^{+-}$  &   38.9         &   5.5$\cdot10^{-5}$     \\
 $1^{+-}$  &   39.8         &   3.8$\cdot10^{-5}$     \\
 $2^{--}$  &   39.8         &   3.8$\cdot10^{-5}$     \\
 $3^{+-}$  &   39.8         &   3.8$\cdot10^{-5}$     \\
 $3^{--}$  &   41.0         &   2.4$\cdot10^{-5}$     \\
 $2^{++}$  &   43.0         &   1.1$\cdot10^{-5}$     \\
 $1^{-+}$  &   45.4         &   4.1$\cdot10^{-6}$     \\
 $0^{-+}$  &  103.6         &   3.5$\cdot10^{-17}$    \\
 $0^{+-}$  &  129.2         &   $\le$1$\cdot10^{-20}$ \\
 $0^{++}$  &  163.1         &   $\le$1$\cdot10^{-20}$ \\
\botrule
\end{tabular}}
\end{table}

\begin{figure}
\centerline{\psfig{file=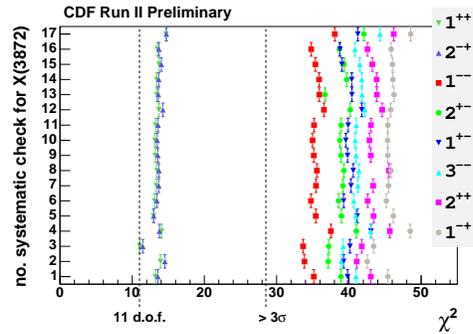,width=2.6in}}
\caption{Variation of the total $\chi^2$ for different
analysis variation. Vertical error bars are for visual guidance.}
\label{fig4}
\end{figure}
To evaluate the stability of the result, we investigate several
effects. The result of the investigation is shown in Figure
\ref{fig4}, where the x-axis shows the resulting $\chi^2$, while
the y-axis denotes the studied effect. 
The default analysis is shown as variation {\em 1}. We
investigate the following variations: {\em 2,3\/} variation in
the fit window, {\em 4,5\/} variation of the bin width, {\em
6,7\/} variation of the Gaussian position, {\em 8,9\/} 
variation of the Gaussian width, {\em 10-12\/} variations in
the dipion mass distribution, {\em 13,14\/} variation of the
$p_T$ and $\eta$ distribution of the $X(3872)$ and {\em
15-17\/} variation of the details of acceptance correction. 
From 
Figure \ref{fig4} we conclude,
that none of the studied effects can alter the conclusion
of the analysis.

\section{Interpretation and Conclusions}

After constraining the quantum numbers of the $X(3872)$ we 
come back to the question of the nature of $X(3872)$. The
natural explanation is that the $X(3872)$ is a conventional
charmonium state. In this picture the state with
$J^{PC}=1^{++}$ could be identified with $\chi_{c1}^{\prime}$ and
the $J^{PC}=2^{-+}$ with the $1^1D_2$ state. However in both
cases there is some difficulty with the conventional
explanation as the predicted
masses\cite{ccMass} are different than the observed value. In addition
the decay to $J/\psi\,\rho$ would violate isospin. But
these arguments alone are not enough to rule out conventional
charmonium. 

The curious fact that the mass is close to the
$D^0\overline{D}^{*0}$
threshold gives raise to the speculations about exotic
interpretation of the $X(3872)$. The most popular exotic
interpretation is that it is $D^0\overline{D}^{*0}$
molecule. 
The idea of the molecular interpretation dates back to mid
seventies \cite{rujula}.
Recently the models of a molecular state
were developed by Tornqvist\cite{Tornqvist} and
Swanson\cite{Swanson}. For a molecular state they predict the
quantum numbers $J^{PC}=1^{++}$, which is compatible with
the result of the analysis. 

It should be added, that the recent observation of the
$X(3872)\rightarrow D^0\overline{D}^0\pi^0$ by
Belle\cite{BelleDDPi} prefers the
$1^{++}$ assignment compared to the $2^{-+}$.

To conclude we presented a determination of the $J^{PC}$  quantum
numbers of the $X(3872)$ state using dipion invariant mass
distribution and angular analysis. We find,
that only the assignments $1^{++}$ and $2^{-+}$ are able to
describe data. All other tested assignments are excluded 
by more than $3$ sigma. While this
result significantly constrains $J^{PC}$ both the conventional
charmonium explanation and the exotic one are still
viable options. To distinguish them further studies
both from experiment and theory side are needed.

\section*{Acknowledgements} 
This work was partially supported by the European Community's Human
Potential Programme under contract HPRN-CT-2002-00292.

\end{document}